\documentclass[twocolumn]{aastex61}
\usepackage{CJK}

\hypersetup{linkcolor=red,citecolor=blue,filecolor=cyan,urlcolor=blue}

\shorttitle{CO in the most luminous QSOs}
\shortauthors{Fan et al.}
\begin{document}
\begin{CJK*}{UTF8}{gbsn}

\title{ALMA detections of CO emission in the most luminous, heavily dust-obscured quasars at $z>3$}

\correspondingauthor{Lulu Fan; Kirsten K. Knudsen}
\email{llfan.sdu@gmail.com; kirsten.knudsen@chalmers.se}

\author[0000-0003-4200-4432]{Lulu Fan (范璐璐)}
\affil{Shandong Provincial Key Lab of Optical Astronomy and Solar-Terrestrial Environment, Institute of Space Science, Shandong University, Weihai, 264209, China}

\author[0000-0002-7821-8873]{Kirsten K. Knudsen}
\affiliation{Department of Space, Earth and Environment, Chalmers University of Technology, Onsala Space Observatory, SE-439 92 Onsala, Sweden}

\author[0000-0003-0003-000X]{Judit Fogasy}
\affiliation{Department of Space, Earth and Environment, Chalmers University of Technology, Onsala Space Observatory, SE-439 92 Onsala, Sweden}

\author[0000-0003-2275-5466]{Guillaume Drouart}
\affiliation{International Centre for Radio Astronomy Research, Curtin University, Perth, Australia}

\begin{abstract}

We report the results of a pilot study of CO$(4-3)$ emission line of three {\it WISE}-selected hyper-luminous, dust-obscured quasars (QSOs) with sensitive ALMA Band 3 observations. These obscured QSOs with $L_{\rm bol}>10^{14}L_\odot$ are among the most luminous objects in the universe. All three QSO hosts are clearly detected both in continuum and in CO$(4-3)$ emission line. Based on CO$(4-3)$ emission line detection, we derive the molecular gas masses ($\sim 10^{10-11}$ M$_\odot$), suggesting that these QSOs are gas-rich systems. We find that three obscured QSOs in our sample follow the similar $L'_{\rm CO}- L_{\rm FIR}$ relation as unobscured QSOs at high redshifts. We also find the complex velocity structures of CO$(4-3)$ emission line, which provide the possible evidence for gas-rich merger in W0149+2350 and possible molecular outflow in W0220+0137 and W0410$-$0913. Massive molecular outflow can blow away the obscured interstellar medium (ISM) and make obscured QSOs evolve towards the UV/optical bright, unobscured phase. Our result is consistent with the popular AGN feedback scenario involving the co-evolution between the SMBH and host galaxy.

\end{abstract}


\section{Introduction}

Quasi-Stellar Objects (QSOs), among the most important astronomical discoveries in the 1960s \citep{schmidt1963}, are believed to be powered by the central supermassive black holes (SMBHs), which accrete gas from a surrounding disk with gravitational energy converted into kinetic energy. The relatively short duty-cycle timescale of 10-100 Myr \citep[e.g.,][]{shen2007} suggests that a significant amount of gas should be available in the proximity of SMBH, such that it can be captured. However, it is not well understood yet which physical process is mainly responsible to efficiently strip angular momentum of gas and transport it from galactic scale ($\sim$kpc) to accretion disk within the central few parsecs.

For massive galaxies and QSOs, the most popular scenario involving the co-evolution between the SMBH and host galaxy proposes that a gas-rich major merger can funnel gas into the galaxy center, triggering the central starburst and feeding the SMBH accretion \citep[e.g.,][]{sanders1988,alexander2012,kormendy2013}. In this scenario, star formation has eventually been quenched by QSO feedback, which is able to heat and expel the ambient gas \citep{dimatteo2005,fabian2012}.

However, observational evidence that supports this merger-driven feeding mechanism over others, such as violent disk instabilities, colliding clouds, or supernova explosions \citep[e.g.,][]{jogee2006}, has remained elusive. Based on optical morphological studies of Active Galactic Nucleus (AGN) host galaxies, the X-ray selected AGN hosts do not show a higher merger fraction than non-active galaxies, both at $z<1$ \citep[e.g.,][]{cisternas2011} and at $z\sim2$ \citep[e.g.,][]{schawinski2011,fan2014,rosario2015}. For luminous QSOs, a high merger fraction ($\sim60-90\%$) has been found \citep{urrutia2008,glikman2015,fan2016a}. This may lead to an explanation that the merger fraction is dependent on the AGN bolometric luminosity \citep[e.g.,][]{treister2012,fan2016a}, which is consistent with theoretical studies, suggesting that galaxy mergers only trigger luminous AGN activity \citep[e.g.,][]{hopkins2009}. 

Recently, massive molecular gas outflow on galactic scales has been observed in luminous QSOs at both low and high redshifts \citep[e.g.,][]{maiolino2012,cicone2014,feruglio2017}.
Moreover, these powerful outflows have been proven to be able to act as negative QSO feedback and affect their host galaxies and suppress star formation in the regions impacted by the outflows \citep[e.g.,][]{carniani2017}.

Given the central role of cold gas in the QSO feeding and feedback processes, it is crucial to trace molecular gas in QSO hosts in order to test the different SMBH-host co-evolution scenarios. Taking the advantage of sensitive sub-millimeter interferometric arrays, the observational studies of the CO emission lines can provide estimates of the total amount of gas available to fuel starburst and AGN activity, and estimates of the galaxy kinematics, such as dynamical mass and/or size of the emitting region \citep{solomon2005}. 

In this work, we present sensitive  ALMA CO$(4-3)$ observations of three hyper-luminous QSOs, taken from a {\it WISE}-selected, heavily dust-obscured sample \citep{wu2012,eisenhardt2012}. Obscured quasars could represent the critical transition phase between starburst and unobscured QSO activity. Our purpose is to study the molecular gas properties and search the clues for feeding and feedback processes in the most luminous obscured QSOs. Throughout this work we assume a flat ${\rm \Lambda}$CDM cosmology with $H_0 = 70$ km~s$^{-1}$, $\Omega_M = 0.3$, and $\Omega_\Lambda = 0.7$.

\section{Targets, Observations and Analysis}

\begin{figure*}
\includegraphics[width=0.95\textwidth]{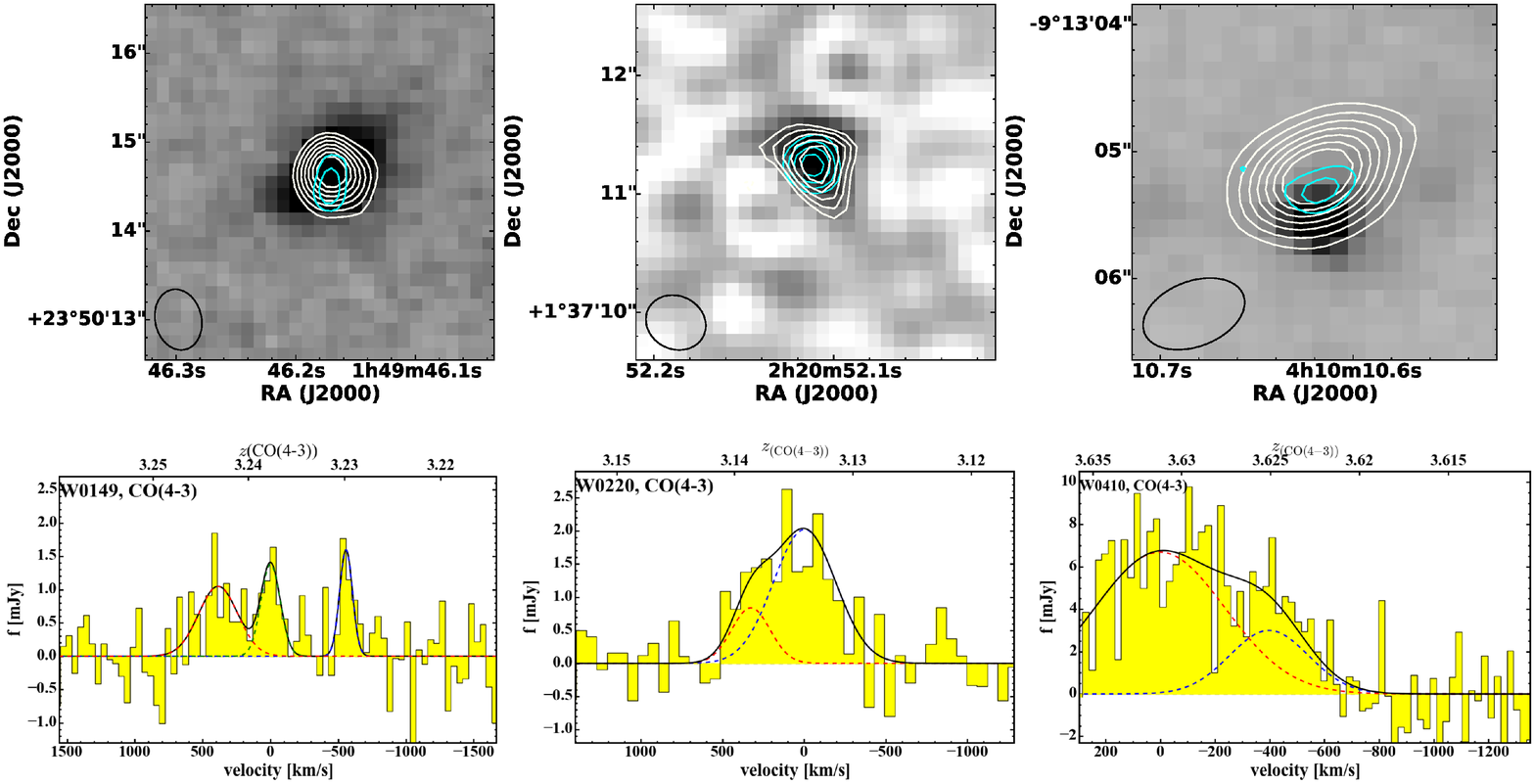}
\caption{{\it Top panel:} Multiwavelength ALMA and {\it HST} images of hyper-luminous, heavily dust-obscured QSOs (from left to right: W0149+2350, W0220+0137 and W0410$-$0913). For W0149+2350 and W0410$-$0913, we plot ALMA CO$(4-3)$ emission line (white) and dust continuum (cyan) contour maps overlaid on the {\it HST} F160W image. For W0220+0137, we plot ALMA CO$(4-3)$ emission line (white) and dust continuum (cyan) contour maps overlaid on the dust continuum image, as {\it HST} F160W image is not available for this object. White contour levels step in units of $(3+2i)\sigma$. Cyan lines show the dust continuum in the 3, 4 and 5$\sigma$ levels. All line maps are continuum subtracted. {\it Bottom panel:} The ALMA CO$(4-3)$ line detections for W0149+2350 (left), W0220+0137 (middle) and W0410$-$0913 (right). Continuum emission has been subtracted. For W0149+2350, the overlaid and color-coded curves represent the Gaussian line fitting for three line components, shown with dashed blue, green and red lines. For W0220+0137 and W0410$-$0913, the lines represent the best-fits obtained with a double Gaussian (each component is shown with dashed red and blue lines). The black solid line shows the sum of the Gaussian curves. 
\label{fig:coimg}}
\end{figure*}

\begin{table*}
\begin{center}
\caption{Summary of the ALMA observations.  
\label{tab:obssum} }
\begin{tabular}{ccccccccc}
\hline
\hline
Source$\, ^{(1)}$ & $z_{opt}\, ^{(2)}$ &  Date$\, ^{(3)}$ & $N_{\rm ant}\, ^{(4)}$ & \multicolumn{3}{c}{Calibrators$\, ^{(5)}$} & $\nu_{\rm spw, central}\, ^{(6)}$ & $\nu_{\rm cont, central}\, ^{(7)}$ \\
 &  & [dd-mm-yyyy] & & Bandpass & Flux & Gain & [GHz] & [GHz] \\
\hline
W0149+2350   & 3.228  & 03-08-2016 & 40 & J0237+2848 & J0238+1636 & J0151+2517 & 109.054 & 97.178, 99.008, 110.886 \\
W0220+0137   & 3.122  & 25-07-2016 & 44 & J0238+1636 & J0006$-$0623 & J0219+0120 & 109.998 & 97.915, 99.806, 111.856 \\
W0410$-$0913 & 3.592  & 24-07-2016 & 39 & J0423$-$0120 & J0423$-$0120 & J0407$-$1211 & 100.393 & 90.310, 88.451, 102.252 \\
\hline
\end{tabular}
\parbox{180mm} {
\textbf{Notes:} \\
(1) The name of the ALMA-observed dust-obscured QSO; (2) Optical redshift; (3) Date of observations ; (4) Number of antennas ($N_{\rm ant}$); (5) Three columns of calibrators used for each set of observations; (6) The central frequency of spectral window used for the line observations; (7) The central frequencies of the three spectral windows used for continuum observations.
\\
}
\end{center}
\end{table*}


We select three hyper-luminous, heavily dust-obscured QSOs (W0149+2350, W0220+0137 and W0410$-$0913) from a new population recently discovered in \textit{Wide-field Infrared Survey Explorer} ({\it WISE}, \citealt{wright2010}) all-sky survey, by using a so-called ``{\it W1W2} dropout" method \citep{wu2012,eisenhardt2012}. The criterion is to select objects which are prominent in the {\it WISE} 12 $\mu m$ ({\it W3}) or 22 $\mu m$ ({\it W4}) bands, and faint or undetected in the 3.4 $\mu m$ ({\it W1}) and 4.6 $\mu m$ ({\it W2}) bands. This selection is very effective for finding high luminosity dust-obscured galaxies with dominant hot dust emission mostly at redshift $z\sim 1-4$ \citep{assef2015a}.  Using X-ray observations \citep{stern2014,piconcelli2015,ricci2017} and the detailed spectral energy distribution (SED) analysis \citep{fan2016b}, clear evidence has been found that the selected mid-IR luminous objects are actually highly dust-obscured, possibly Compton-thick AGN, with a high accretion rate close to the Eddington limit \citep{wu2018}.   

Observations were carried out with ALMA during Cycle 3 using the Band 3 receiver. Summary of observations is given in Table \ref{tab:obssum}. For each source, the receiver was tuned to the redshifted CO$(4-3)$ line using the optical redshift taken from \citet{wu2012}. The spectral window of the redshifted CO$(4-3)$ line used a setup for spectral line mode, while the three remaining spectral windows used a continuum mode setup. The telescope configuration included baselines between 15 and 1124\,m (though 1396\,m for W0149+2350).  Reduction, calibration, and imaging were done using CASA (Common Astronomy Software Application\footnote{https://casa.nrao.edu}; \citealt{mcmullin07}). The results from the pipeline reduction\footnote{For details on the pipeline, see https://almascience.eso.org/documents-and-tools/} carried out by the observatory were generally sufficient with only some minor extra flagging, which did not significantly change the final result.  For the absolute flux calibration, we adopted a conservative uncertainty of 10\%.

\begin{table*}
\begin{center}
\caption{Properties of three hyper-luminous, heavily dust-obscured QSOs at $z=3.1-3.6$
\label{tab:almares}}
\begin{tabular}{lcccccccccc}
\hline
\hline
Source & RA(J2000) & Dec(J2000) & $z_{\rm CO(4-3)}$ & $S_{\rm peak}$ & FWHM & $I_{\rm CO(4-3)}$ &  $L'_{\rm CO(4-3)}$ & $M_{\rm gas}$ \\
 & hhmmss.s & ddmmss.s & & [mJy] & [km\,s$^{-1}$] & [Jy\,km\,s$^{-1}$] & [$10^{10}$\,K\,km\,s$^{-1}$\,pc$^2$] & [$10^{10}\,M_\odot$] \\
\hline
W0149+2350  & \\
W0149-comp1    & 01:49:46.17 &  +23:50:14.66 & $3.2432\pm0.0005$ & $1.05\pm0.23$ & $340\pm 98$ & $0.38\pm0.14$ & $1.12\pm0.40$ & $0.8\pm0.3$ \\
W0149-comp2    & 01:49:46.17 &  +23:50:14.66 & $3.2377\pm0.0003$ & $1.38\pm0.31$ & $167\pm 48$ & $0.25\pm0.09$ & $0.72\pm0.26$ & $0.5\pm0.2$ \\
W0149-comp3    & 01:49:46.17 &  +23:50:14.66 & $3.2298\pm0.0002$ & $1.60\pm0.38$ & $112\pm 31$ & $0.19\pm0.07$ & $0.56\pm0.20$ & $0.4\pm0.1$  \\
W0149-total  &  & & & & & $0.81\pm 0.18$ & $2.40\pm0.50$  & $1.7\pm0.4$ \\ 
W0220+0137  & \\
W0220-comp1   & 02:20:52.124 &  +01:37:11.23 & $3.1341\pm0.0004$ & $2.02\pm0.28$ & $449\pm 73$ & $0.97\pm0.20$ & $2.69\pm0.57$ & $1.9\pm0.4$  \\
W0220-comp2   & 02:20:52.107 &  +01:37:11.22 & $3.1386\pm0.0005$ & $0.84\pm0.24$ & $277\pm 94$ & $0.25\pm0.11$ & $0.69\pm0.30$ & $0.5\pm0.2$ \\ 
W0220-total  & & & & & & $1.21\pm 0.23$ & $3.39\pm0.65$ & $2.4\pm0.5$   \\
W0410$-$0913  &  \\
W0410-comp1     & 04:10:10.606 &$-$09:13:04.96 & $3.6312\pm0.0003$ & $6.69\pm0.37$ & $560\pm 44$ & $3.99\pm0.38$ & $14.10\pm1.33$ & $9.9\pm0.9$ & \\ 
W0410-comp2     & 04:10:10.621 &$-$09:13:05.30 & $3.6251\pm0.0003$ & $3.00\pm0.35$ & $338\pm 46$ & $1.08\pm0.19$ & $3.80\pm0.67$ & $2.7\pm0.5$   \\
W0410-total & & & & & & $5.07\pm 0.42$ & $17.90\pm1.50$ &  $12.5\pm1.1$  \\
\hline
\end{tabular}
\end{center}
\end{table*}
\bigskip

\section{Results}

Figure \ref{fig:coimg} presents multiwavelength ALMA and {\it HST} images and the ALMA CO$(4-3)$ emission line spectra of three heavily dust-obscured QSOs. All three quasar hosts are clearly detected both in dust continuum and in CO$(4-3)$ line emission. Using UVMULTIFIT \citep{martividal14}, we estimate the continuum and emission line properties. The CO line luminosities are derived from the Gaussian fit results, using Equation 3 in \citet{solomon2005}. For each CO$(4-3)$ emission line component, the redshift has been estimated based on the Gaussian fit. All lines are assumed to be Gaussian and the results are summarized in Table~\ref{tab:almares}. 

{\it W0149+2350} : 
We plot the integrated  CO$(4-3)$ emission line (white) and dust continuum (cyan) contour maps overlaid on the {\it HST} F160W image. In the line spectral window, the rms is $\sim0.59$\,mJy/beam per 21.5\,km\,s$^{-1}$ channel. Three CO line components are detected together with a continuum detection at 4.8$\sigma$ level. The velocity separations of two adjacent line components are 380 and 560\,km\,s$^{-1}$, respectively. The angular resolution is 0.2\,arcsec, which corresponds to $\sim$1.5\,kpc at $z\sim3.2$. At the scales of the observations, it is not possible to estimate the angular separation between the different line components and also determine if the continuum emission arises from one or more components. The angular extension is consistent with a point source, though there is a minor indication that the emission could be extended. 

{\it W0220+0137} : 
We plot the integrated  CO$(4-3)$ emission line (white) and dust continuum (cyan) contour maps overlaid on the dust continuum image, as {\it HST} F160W image is not available for this object. In the line spectral window, the rms is $\sim1.04$\,mJy/beam per 21.1\,km\,s$^{-1}$ channel. We fit the CO$(4-3)$ line with a double Gaussian, which gives a better fit with the reduced $\chi^2=0.74$ than a single Gaussian does with reduced $\chi^2=0.86$. The two emission line components have a $\sim 320$\,km\,s$^{-1}$ velocity separation and an angular separation of 0.25\,arcsec, which corresponds to 1.9\,kpc at $z\sim3.1$. We can not determine which emission line component is associated with the continuum emission. The fainter emission component appears to be marginally extended by $0.39\pm0.1$\,arcsec, while the brighter CO line component is unresolved.  

{\it W0410$-$0913} : 
We plot the integrated  CO$(4-3)$ emission line (white) and dust continuum (cyan) contour maps overlaid on the {\it HST} F160W image. In the line spectral window, the rms is $\sim0.84$\,mJy/beam per 23.5\,km\,s$^{-1}$ channel. The optical redshift of W0410$-$0913 presented in \citet{wu2012} was based on relatively faint C\,{\sc iv} and He\,{\sc ii} lines and no formal error bar was given.  The CO line is detected, though very close to the low-frequency edge of the band, which corresponds to a shift of $\sim 2500$\,km\,s$^{-1}$ from the optical redshift.  We fit the CO$(4-3)$ line with a double Gaussian, which gives a better fit with the reduced $\chi^2=1.15$ than a single Gaussian does with reduced $\chi^2=1.32$. Two CO line components have a velocity separation of 395\,km\,s$^{-1}$ and an angular separation of 0.4\,arcsec, which corresponds to 3.2\,kpc.  The brighter CO component is extended measuring FWHM\,=\,$0.53\pm0.04$\,arcsec for a 2D circular Gaussian function, while the fainter CO component is unresolved.  The continuum emission is coincident with the bright CO component.  As the CO emission is detected close to the edge, the red tail of the bright, broad emission is truncated. Thus we cannot exclude if a third component is also presented on that side. 

Given the relatively large frequency coverage of the observations, the continuum measurements also reveal a change in flux density from upper to lower sidebands. For W0149+2350, W0220+0137 and W0410$-$0913, we measure the continuum emissions at 98.09/109.97 GHz (lower/upper sideband), 98.86/110.93 GHz and 89.38/101.32 GHz, respectively. The corresponding flux densities are $111\pm28$/$171\pm41$ $\mu$Jy, $78\pm24$/$180\pm31$ $\mu$Jy and $121\pm37$/$258\pm40$ $\mu$Jy.    

\begin{figure}
\epsscale{1.25}
\plotone{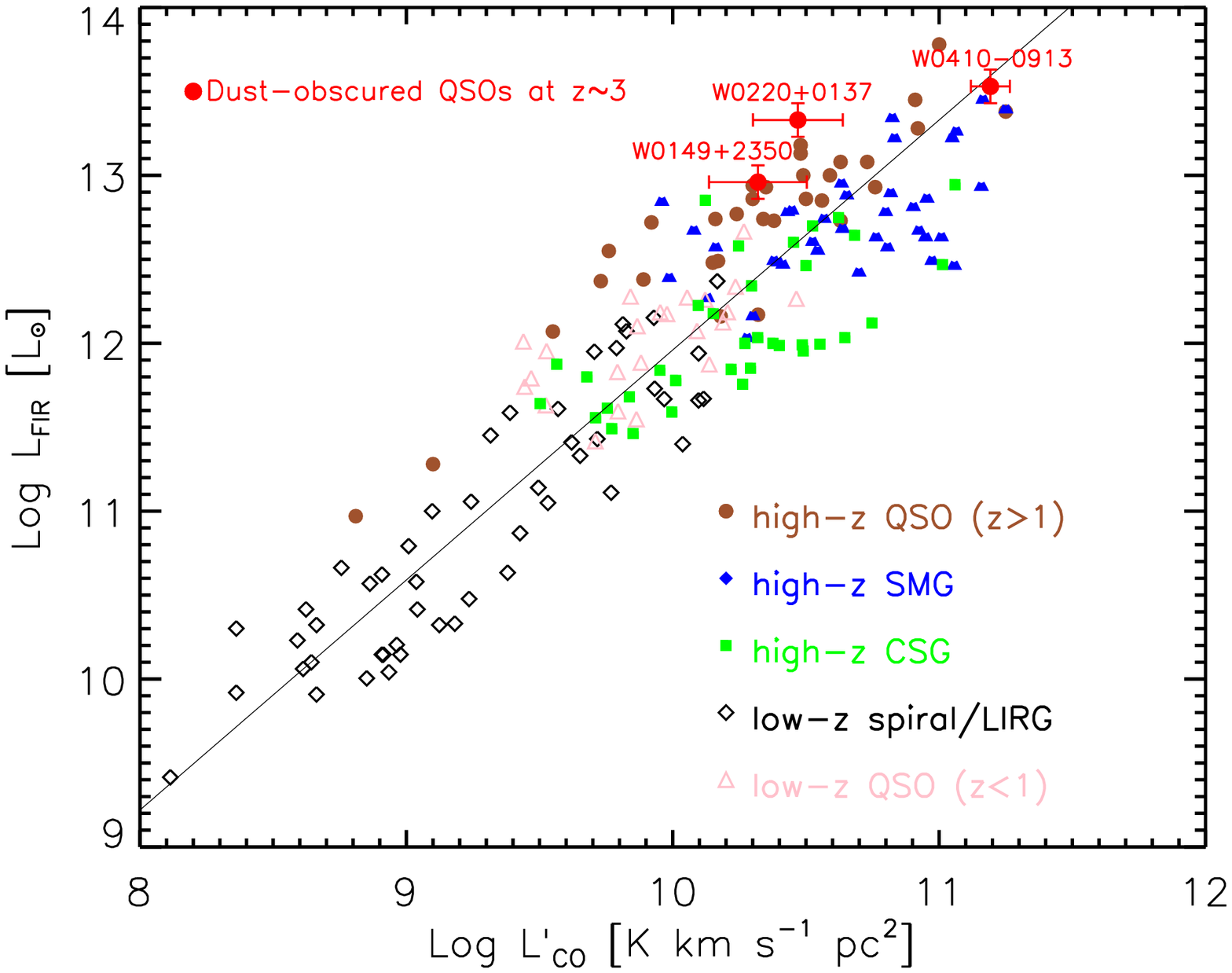}
\caption{The correlation between $L'_{\rm CO}$ and $L_{\rm FIR}$ for high-redshift QSOs, submillimeter galaxies (SMGs) and color-selected galaxies (CSGs) taken from the compilation in \citet{carilli2013}, together with nearby spiral, luminous infrared galaxies \citep[LIRGs;][]{gao2004}, low-redshift QSOs \citep{xia2012} and  three hyper-luminous, heavily dust-obscured QSOs (this work). The solid line corresponds to a fit result $log\ L_{\rm FIR}=1.37\ log\ L'_{\rm CO}-1.74$ with all data points. 
\label{fig:lcolfir}}
\end{figure}

\section{Discussions}

\begin{figure*}
\centering
\includegraphics[width=0.95\textwidth]{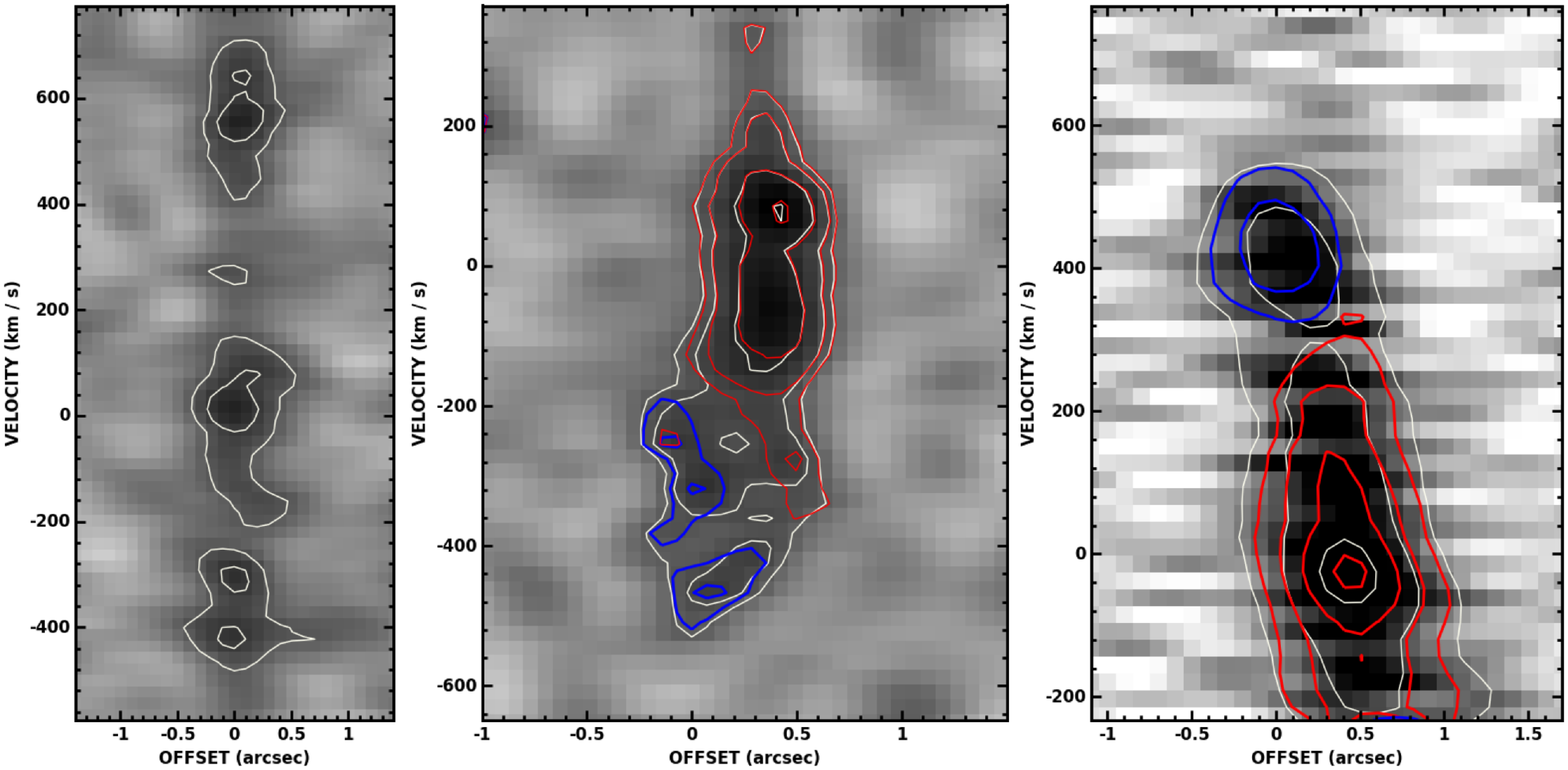}
\caption{Position-velocity (PV) diagrams of the CO$(4-3)$ emission line for W0149+2350 (left), W0220+0137 (middle) and W0410$-$0913 (right). The PV diagrams have been extracted along PA = 0, 73 and $-30$ deg, respectively. For W0220+0137 and W0410$-$0913, red and blue contours correspond to two CO$(4-3)$ emission line components in Figure \ref{fig:coimg}.
\label{fig:pv}}
\end{figure*}

As shown in Figure \ref{fig:coimg}, our ALMA Band 3 observations have successfully detected the CO$(4-3)$ line emission of three hyper-luminous, heavily dust-obscured QSOs at $z>3$. Adopting QSO excitation $L'_{\rm CO(4-3)}/L'_{\rm CO(1-0)} = 0.87$ and the CO-to-gas mass conversion factor $\alpha_{CO}=0.8 M_\odot\,({\rm K\,km\,s^{-1}\,pc^2})^{-1}$ \citep{carilli2013}, we derive the molecular gas masses of the three dust-obscured QSOs (see Table \ref{tab:almares}). The total gas masses range from $10^{10}$ to $10^{11}$ M$_\odot$, suggesting that these QSOs are gas-rich systems. 

We estimate the far-infrared (FIR) luminosity $L_{\rm FIR}$ of three hyper-luminous, heavily dust-obscured QSOs with SED decomposition method used in our previous work \citep{fan2016b}. The new ALMA continuum measurements have been included. However, we take the estimated $L_{\rm FIR}$ as the {\it total} luminosity of different components considering the blended {\it Herschel} photometry. Except for W0149+2350, a larger $\beta$ (2.0, instead of the previously adopted value 1.6) is required to fit the new IR SEDs of W0220+0137 and W0410$-$0913  which extend to the observed-frame 3 mm. Despite adopting a different $\beta$, the derived $L_{\rm FIR}$ only have a slight difference (up to 0.1 dex). In Figure \ref{fig:lcolfir}, we plot the correlation between $L'_{\rm CO}$ and $L_{\rm FIR}$ for the three dust-obscured QSOs and other populations from compilation in \citet{carilli2013}. We find that the three dust-obscured QSOs follow the similar $L'_{\rm CO}- L_{\rm FIR}$ relation as unobscured QSOs at high redshifts.

In the top panel of Figure \ref{fig:coimg}, the velocity-integrated CO emission line map (white contour) shows a single line component for each QSO. However, we emphasize that complexity is revealed by the CO emission line spectra in the bottom panel of Figure \ref{fig:coimg}. For W0149+2350, it is clear there are three line components. For W0220+0137 and W0410$-$0913, a double Gaussian model provides a better fit than a single Gaussian one, despite the choice is not as convincing as for W0149+2350 due to the limited signal-to-noise. In Figure \ref{fig:pv}, we plot position-velocity (PV) diagrams of the CO$(4-3)$ emission line for W0149+2350, W0220+0137 and W0410$-$0913. The PV-diagrams also suggest that all of three QSOs have complex velocity structures, which may disfavor the scenario of rotation disk. W0149+2350, showing three distinguishable line components, is possibly undergoing a gas-rich merger, though the scenario of having a molecular outflow can not be excluded. W0220+0137 and W0410$-$0913 have asymmetric velocity structures, suggesting possible molecular outflows, though a gas-rich merger scenario could also explain this. The possibility that they are just a single Gaussian component can not be fully excluded. For W0410$-$0913, the red side of the line has been truncated as the CO line is detected close to the edge of the spectral window.  

Recently, direct measurements of the gas content of high-redshift QSOs start to be available for a large number of sources thanks to the observations of CO and [CII] lines with ALMA. While the QSO host properties are not accessible in the case of Type 1 QSOs (unobscured) --- and therefore a quantitative mass-selected comparison is not possible yet --- these systems appear to have a similar molecular gas properties to allow us to draw some qualitative comparisons with our sample. In particular, the sample from \citet{banerji2017}, with obscured QSOs at $z\sim3$ is relatively close to our sample. Their sample only presents a single CO component, centered on the continuum emission, leading to believe that close-merger pairs are absent ($<10$\,kpc). As a comparison, one out of our three obscured QSOs (W0149+2350) is possibly undergoing an advanced gas-rich merger. However, due to the small sample size, the difference is not significant. 

Previous JCMT SCUBA-2 $850\, \mu m$ follow-up observations have detected overdensities of companion SMGs around these {\it WISE}-selected obscured QSOs on scales of several hundreds kpc \citep{jones2014,jones2015,fan2017}. \citet{banerji2017} noted the presence of star forming companions in the vicinity of some QSOs (100\,kpc-scale) which provided evidence for dense environments. For the samples at even higher redshifts ($4.8<z<7$), studies of Type 1 QSOs \citep{wang2013,trakhtenbrot2017,venemans2017} also revealed companions associated with a significant fraction of QSOs at separations at the 10s\,kpc-scale, as well as showing a single molecular gas component associated with the continuum emission. In general, finding companion galaxies around QSOs is expected from theoretical modeling \citep[e.g.][]{fogasy2017}. However, we have not found any serendipitous detection of dust continuum or CO emission line around our obscured QSOs within ALMA Band 3 field of view (FOV, $\sim100$ kpc at $z\sim3$). The difference between our result and others may be due to several aspects, such as the small sample size, the different FOV, observed wavelength and detection sensitivity. 

At least two out of three obscured QSOs (W0220+0137 and W0410$-$0913) in our sample show possible evidence for galactic-scale molecular outflow. This finding is consistent with the recent [CII] studies of another hyper-luminous obscured QSO, W2246$-$0526, selected from the same sample as ours \citep{diaz2016}. They found that W2246$-$0526 is blowing out its ISM isotropically in a homogeneous, large-scale turbulent outflow. Similar molecular outflows have also been found in other high-redshift obscured QSOs \citep[e.g.,][]{polletta2011,brusa2017}, suggesting that it may be a common feature of obscured QSOs. Galactic-scale molecular outflow can play an important role, which can blow away the obscured ISM and make the obscured QSO evolve to be UV/optical bright and unobscured. This scenario is consistent with the expectation of SMBH-host co-evolution model \citep{sanders1988,dimatteo2005,alexander2012}.

\section{Summary}

In this letter, we present a pilot study of ALMA observations of CO$(4-3)$ emission line in three {\it WISE-}selected, heavily dust-obscured QSOs. For all three obscured QSOs, we clearly detect both continuum and CO$(4-3)$ emission line. Based on CO$(4-3)$ line detection, we derive the total molecular gas masses ranging from $10^{10}$ to $10^{11}$ M$_\odot$. Given the high FIR luminosity and CO$(4-3)$ line luminosity, the three obscured QSOs follow the similar $L'_{\rm CO}- L_{\rm FIR}$ relation as unobscured QSOs at high redshifts, indicating that they are possibly similar objects in the different evolutionary stages. All of the CO$(4-3)$ lines show the complex velocity structures. One out of three obscured QSOs (W0149+2350) is possibly undergoing an advanced gas-rich merger. The other two obscured QSOs (W0220+0137 and W0410$-$0913) have possible molecular outflows, which is consistent with the AGN feedback scenario. Our obscured QSOs may represent a brief evolutionary stage before obscured ISM have been cleared, evolving into UV/optical bright QSOs.

\acknowledgments

We thank the anonymous referee for his/her comments and suggestions, which have greatly improved this paper. We thank the staff of the Nordic ALMA Regional Center node for their support and helpful discussions. This work is supported by National Key R\&D Program of China (No. 2017YFA0402703). LF acknowledges the support from the National Natural Science Foundation of China (NSFC, Grant Nos. 11773020 and 11433005) and Shandong Provincial Natural Science Foundation, China (ZR2017QA001). KK and JF acknowledge the Knut and Alice Wallenberg Foundation for support. 

This paper makes use of the following ALMA data: ADS/JAO.ALMA\#2015.1.01148.S (PI: Lulu Fan). ALMA is a partnership of ESO (representing its member states), NSF (USA) and NINS (Japan), together with NRC (Canada), MOST and ASIAA (Taiwan), and KASI (Republic of Korea), in cooperation with the Republic of Chile. The Joint ALMA Observatory is operated by ESO, AUI/NRAO and NAOJ.

\textit{Facilities}: ALMA

\end{CJK*}
\end{document}